\documentclass[referee]{aa} 
\usepackage{graphicx}
\usepackage{txfonts}
\begin{document}
\title{Advection-Dominated Accretion Flows with Causal Viscosity}
\author{R. Takahashi\inst{1}}
\offprints{R. Takahashi}

\institute{
Graduate School of Arts and Sciences, University of Tokyo, 153-8902, Tokyo, Japan\\
\email{rohta@provence.c.u-tokyo.ac.jp}
}

\date{Received August XX, 200X; accepted March XX, 200X}
 
\abstract 
{}
{
We present the basic equations and sample solutions for 
the steady-state global transonic solutions of the advection-dominated accretion flows (ADAFs) 
with causal viscosity prescription. 
The procedures of the stable numerical calculations and all explicit formula 
to obtain the solutions of ADAFs are also presented. 
}
{
We solve the transonic solutions of ADAFs by using the explicit numerical integrations, 
such as the Runge-Kutta method. 
In this calculation method, we firstly solve the physical values at the sonic radius 
where L'Hopital's rule is used. 
Then, we numerically solve the coupled differential equations of the radial velocity, 
the angular momentum and the sound speed from the sonic radius 
to the inward and outward directions.  
}
{%
By the calculation procedures presented in this paper, 
we can cover the all parameter spaces of the transonic solutions of ADAFs.  
Sample transonic solutions for ADAF-thick disk and ADAF-thin disk are presented. 
The explicit formula for 
the analytical expansion around the singular points, the sonic point and the viscous point, 
are presented. 
If we set the diffusion timescale to be null, 
the formalism in this study become the formalism of the acausal viscosity which is usually 
used in the past study for the calculation of the ADAF structure. 
}
{}


\maketitle
%

\section{Introduction}

Advection dominated accretion flows (ADAFs) are one of the fundamental solutions of 
the accretion flows into the black holes in the low-luminous systems, such as, 
low-luminosity active galactic nuclei. 
In past studies, the several methods are used to obtain the transonic solutions of ADAFs, e.g., 
the relaxation scheme (e.g. Narayan, Kato \& Honma 1997), 
the implicit integration scheme (e.g. Kato, Fukue \& Mineshige 1998),  
the outward integration scheme with the inner boundary condition (Becker \& Le 2003) 
or the inward or outward integration scheme with the special condition 
at and inside the sonic radius (Lu, Gu \& Yuan 1999). 
All these works assumes the diffusive viscosity with infinite diffusive velocity. 
In this paper, we give the method to solve the global transonic solutions of ADAFs with 
causal viscosity prescriptions which assume the finite diffusive velocity proposed by 
Papaloizou \& Szuszkiewicz (1994), which can become the alternative method to solve the 
global solutions.  
Gammie \& Popham (1998) calculate the ADAF structure in the Kerr metric with the 
causal viscousity prescription by using the relaxation scheme. 
In the calculations with the causal viscosity prescription, 
the boundary condition at the viscous point is used, and the boundary condition 
at the horizon is not required (see, also \S 5). 
If we set the diffusion timescale to be null, 
the basic equations in this study become the basic equations for the case of 
the acausal viscosity which is usually used in the past study. 
After we present the basic equations for the ADAF with the causal viscosity prescription in \S 2, 
the boundary conditions for the global transonic solutions of ADAFs with the causal viscosity 
are given in \S 3. 
The numerical procedures of the method to solve the transonic solutions of ADAFs 
and the sample solutions are presented in \S 4. 
Discussion and conclusions are given in \S 5 and \S 6, respectively. 
%

\section{Basic Equations of ADAF with Causal Viscosity}
The basic equations used in  this paper is same as those in Narayan, Kato \& Honma (1997) except that 
we use the causal viscosity prescription. 
The mass conservation is written as, 
\begin{equation}
-4 \pi r H \rho v_r = \dot{M}, 
\label{eq:Mdot}
\end{equation}
where $H$ is the half thickness of the accretion disk, 
$\rho$ is the gas density, $v_r$ is the radial velocity and 
$\dot{M}$ is the mass accretion rate. 
Here, we assume $v_r<0$ and $H=(5/2)^{1/2}a_s/\Omega_K$ where $a_s$ is the sound speed defined as 
$a_s=(p/\rho)^{1/2}$ and $p$ is the pressure. 
The radial momentum conservation is described as, 
\begin{equation}
v_r \frac{dv_r}{dr}=(\Omega^2-\Omega_K^2)r-\frac{1}{\rho}\frac{dp}{dr}, 
\label{eq:radialv}
\end{equation}
where $\Omega$ is the angular velocity of the accretion flow and $\Omega_K$ is the Keplerian angular 
velocity calculated as 
$\Omega_K=1/[r^{1/2}(r-2)]$. 
Here, we set $GM=1$ and use the Paczy\'{n}skii-Wiita potential. 
The energy equation is written as, 
\begin{equation}
\frac{\rho v_r}{\gamma-1}\frac{da_s^2}{dr}-a_s^2 v_r 
\frac{d\rho}{dr}=\rho v_r (\ell-j)\frac{d\Omega}{dr},
\label{eq:energy}
\end{equation}
where $\gamma$ is the ratio of specific heats of the accreting gas, $\ell(=\Omega r^2)$ is the 
angular momentum, and $j$ is the integration constant meaning 
the specific angular momentum at some radius. 
For the case of the diffusive viscosity with the infinite diffusion velocity, 
the shear stress $t_{r\phi}$ is written as 
\begin{equation}
t_{r\phi}=-\nu \rho r^2\frac{d\Omega}{dr}. 
\label{eq:shear}
\end{equation}
Then, the angular momentum equation for the case of the acausal viscosity is written as, 
\begin{equation}
\frac{d\Omega}{dr}=\frac{v_r(\ell-j)}{\nu r^2}, 
\label{eq:angvel}
\end{equation}
where $\nu$ is the kinematic coefficient of viscosity, and, 
here, $j$ is the specific angular momentum per unit mass where $d\Omega/dr=0$, 
if such point exists.  
By assuming the $\alpha$-viscosity we set $\nu=\alpha a_s^2/\Omega_K$. 
From Eq. (\ref{eq:angvel}), the equation for the angular momentum is written as 
\begin{equation}
\frac{d\ell}{dr}=\frac{2\ell}{r}+\frac{v_r(\ell-j)}{\nu}. 
\label{eq:angmom}
\end{equation}
%

%
In the same way as Papaloizou \& Szuszkiewicz (1994) and Gammie \& Popham (1998), 
we include the effects of the diffusive viscosity with the finite diffusion velocity. 
By using the causal viscosity, we can derive 
the angular momentum equation with the diffusive viscosity with the diffusion velocity $a_\nu$ 
written as, 
\begin{equation}
\frac{d\ell}{dr}=\frac{\mathcal{N}_\nu}{\mathcal{D}_\nu},
\label{eq:angmom2}
\end{equation}
where
\begin{equation}
\mathcal{D}_\nu = 1-v_r^2/a_\nu^2,~~~~
\mathcal{N}_\nu = \frac{2\ell}{r}+\frac{v_r(\ell-j)}{\nu} \left(1-\frac{2v_r \tau_\nu}{r}\right), 
\end{equation}
Here, $\tau_\nu$ is the relaxation timescale. 
For simplicity, we set $\tau_\nu$ and $a_\nu$ as $\tau_\nu=\Omega_K^{-1}$ and 
$a_\nu=(\nu/\tau_\nu)^{1/2}$, respectively. 
Now, the shear stress for the causal viscosity is written as 
%
\begin{equation}
t_{r\phi}=\frac{-\nu \rho}{1-2v_r\tau_\nu/r}
\left[
\left(
1-\frac{v_r^2}{a_\nu^2}
\right)
r^2\frac{d\Omega}{dr}
-\frac{2\ell}{r}\frac{v_r^2}{a_\nu^2}
\right]. 
\label{eq:shear2}
\end{equation}
%
When $\tau_\nu=0$, i.e., $a_\nu=\infty$, 
Eq. (\ref{eq:angmom2}) and (\ref{eq:shear2}) for the causal viscosity 
become the equations for the acausal viscosity, Eq. (\ref{eq:angmom}) and (\ref{eq:shear}). 
Even in the point where $d\Omega/dr=0$, 
the shear stress $t_{r\phi}$ in general do not become null due to effects of the causal viscosity. 
Since, the effects of the causal viscosity are coupled with the radial velocity 
as $v_r^2/a_\nu^2$ and $v_r\tau_\nu$ in Eqs. (\ref{eq:angmom2}) and (\ref{eq:shear2}), 
the effects of the causal viscosity is effective in the inner region of the accretion flow 
where the absolute value of the radial velocity is large. 
%

\section{Boundary Conditions and Physical Values at the Sonic Radius}
Since the accretion flows plunge into the black hole supersonically after a sonic transition, 
the equations are singular at the sonic radius. 
In order to see this, from Eqs. (\ref{eq:Mdot}), (\ref{eq:radialv}) and (\ref{eq:energy}), 
we can derive the equation for $dv_r/dr$ as, 
\begin{equation}
\frac{dv_r}{dr}=\frac{\mathcal{N}}{\mathcal{D}},
\label{eq:dvr}
\end{equation}
where
\begin{eqnarray}
\mathcal{D}&=& v_r-\left(\frac{2\gamma}{\gamma+1}\right)\frac{a_s^2}{v_r},\nonumber \\
\mathcal{N}&=&\frac{\Omega^2-\Omega_K^2}{r}-\left(\frac{2\gamma}{\gamma+1}\right)
				a_s^2 \frac{d\ln (\Omega_K/r)}{dr} 
\nonumber \\
&&
-\left(\frac{\gamma-1}{\gamma+1}\right)\frac{(\ell-j)}{r^2}
	\left(\frac{d\ell}{dr}-\frac{2\ell}{r}\right). 
\end{eqnarray}
Here, Eq. (\ref{eq:angmom2}) is inserted to $d\ell/dr$ and 
$d\ln (\Omega_K/r)/dr=-(5r-6)/[2r(r-2)]$. 
On the other hand, the equation for $da_s/dr$ is written as, 
\begin{eqnarray}
\frac{da_s}{dr}=
\left(\frac{\gamma-1}{\gamma+1}\right)
\left[
-\frac{a_s}{v_r}\frac{dv_r}{dr}+a_s\frac{d\ln (\Omega_K/r)}{dr}
+\frac{(\ell-j)}{a_s r^2}\left(\frac{d\ell}{dr}-\frac{2\ell}{r}\right)
\right].\nonumber\\
\label{eq:das}
\end{eqnarray}

In order to pass the sonic radius, $r_s$, smoothly, two boundary conditions at $r=r_s$ are given as 
$\mathcal{D}=\mathcal{N}=0$ at $r=r_s$. 
For given $a_{s,s}~(>0)$ which is the sound speed at the sonic radius, 
the radial velocity $v_{r,s}$ at $r=r_s$ is calculated from $\mathcal{D}=0$ as
$v_{r,s}=-a_{s,s}[2\gamma/(\gamma+1)]^{1/2}$. 
From the condition $\mathcal{N}=0$, the angular momentum $\ell_s$ at $r=r_s$ is calculated as,  
$\ell_s=[b_\ell\pm(b_\ell^2-a_\ell c_\ell)^{1/2}]/a_\ell$
where 
\begin{eqnarray}
a_\ell &=& 
	1-\left(\frac{\gamma-1}{\gamma+1}\right)
        	\left\{
                -2+\frac{1}{\mathcal{D}_\nu}
                	\left[
                        2+\frac{rv_r}{\nu}
                        	\left(
                                1-\frac{2v_r\tau_r}{r}
                                \right)
                        \right]
                \right\}
	,\\
b_\ell &=&
	\frac{j}{2}\left(\frac{1-\gamma}{1+\gamma}\right)
        \left\{
        	-\frac{r v_r}{\nu\mathcal{D}_\nu}
                \left(
                1-\frac{2v_r\tau_r}{r}
                \right)
        \right. \nonumber\\
        && \left.
        -2+\frac{1}{\mathcal{D}_\nu}
                	\left[
                        2+\frac{rv_r}{\nu}
                        	\left(
                                1-\frac{2v_r\tau_r}{r}
                                \right)
                        \right]
        \right\}
	,\\
c_\ell &=&
	-\left(\frac{\gamma-1}{\gamma+1}\right)\frac{r v_r j^2}{\nu\mathcal{D}_\nu}
        	\left(
                1-\frac{2v_r\tau_r}{r}
                \right)
        -\Omega_K^2 r^4 
        \nonumber\\
        &&-\frac{2\gamma}{\gamma+1}a_s^2 r^3\frac{d\ln (\Omega_K/r)}{dr}
        .
\end{eqnarray}
%

%
Chakrabarti introduce very clever method to calculate the value of $dv_r/dr$ at $r=r_s$ 
(e.g. Chakrabarti 1990, 1996). 
We extend their method to calculate the transonic solutions of the ADAFs with causal viscosity. 
The value of $dv_r/dr$ at $r=r_s$ is calculated by applying L'Hopital's rule as follows. 
To apply L'Hopital's rule to Eq. (\ref{eq:radialv}) at $r=r_s$, we introduce new variables 
$x=r-r_s$, $y=v_r-v_{r,s}$, $z=\ell-\ell_s$ and $w=a_s-a_{s,s}$. 
The functions $\mathcal{D}$ and $\mathcal{N}$ are expanded near the sonic point as, 
%
\begin{eqnarray}
\mathcal{D} &:& 
	\left(\frac{\partial\mathcal{D}}{\partial x}\right)_s x 
        +\left(\frac{\partial\mathcal{D}}{\partial y}\right)_s y 
        +\left(\frac{\partial\mathcal{D}}{\partial z}\right)_s z 
        +\left(\frac{\partial\mathcal{D}}{\partial w}\right)_s w 
        ,\\
\mathcal{N} &:& 
	\left(\frac{\partial\mathcal{N}}{\partial x}\right)_s x 
        +\left(\frac{\partial\mathcal{N}}{\partial y}\right)_s y 
        +\left(\frac{\partial\mathcal{N}}{\partial z}\right)_s z 
        +\left(\frac{\partial\mathcal{N}}{\partial w}\right)_s w 
	, 
\end{eqnarray}
where
\begin{eqnarray}
\left(\frac{\partial\mathcal{D}}{\partial x}\right)_s&=& 
\left(\frac{\partial\mathcal{D}}{\partial z}\right)_s=0,~~~
\left(\frac{\partial\mathcal{D}}{\partial y}\right)_s= 2,~~~
\left(\frac{\partial\mathcal{D}}{\partial w}\right)_s= 2\sqrt{\frac{2\gamma}{\gamma+1}}, 
\nonumber \\
\left(\frac{\partial\mathcal{N}}{\partial x}\right)_s&=& 
	\frac{1}{(r-2)^3}-\frac{3\ell^2}{r^4}
        -\left(\frac{2\gamma}{\gamma+1}\right)a_s^2 \frac{d^2\ln (\Omega_K/r)}{dr^2}
        \nonumber\\
        &&
        -\left(\frac{\gamma-1}{\gamma+1}\right)\frac{6\ell(\ell-j)}{r^4}
	,\nonumber \\
\left(\frac{\partial\mathcal{N}}{\partial y}\right)_s&=& 
	0
	,~~~
\left(\frac{\partial\mathcal{N}}{\partial z}\right)_s= 
	\frac{2\ell}{r^3}
        +\left(\frac{\gamma-1}{\gamma+1}\right)\frac{2(2\ell-j)}{r^3}
	,\nonumber \\
\left(\frac{\partial\mathcal{N}}{\partial w}\right)_s&=& 
	\left( \frac{-4\gamma}{\gamma+1} \right) a_s \frac{d\ln (\Omega_K/r)}{dr}. 
\end{eqnarray}
Here, $d^2\ln (\Omega_K/r)/dr^2=(5r^2-12r+12)/[2r^2(r-2)^2]$, 
and $z$ and $w$ are described as  
$z=(d\ell/dr)_s~x$ and 
$w=(\partial a_s/\partial r)_s~x+(\partial a_s/\partial v_r)_s~y$ 
where 
\begin{eqnarray}
\left(\frac{\partial a_s}{\partial r}\right)_s &=&
	-\left(\frac{\gamma-1}{\gamma+1}\right)\frac{a_s}{v_r}
	, \nonumber \\
\left(\frac{\partial a_s}{\partial v_r}\right)_s &=& 
	\left(\frac{\gamma-1}{\gamma+1}\right)
	\left\{
		a_s \frac{d\ln (\Omega_K/r)}{dr}
	+\frac{(\ell-j)}{a_s r^2}
        	\left[\left(\frac{d\ell}{dr}\right)_s-\frac{2\ell}{r}\right]
	\right\}.\nonumber 
\end{eqnarray}
Finally, we obtain the quadratic equations of $dv_r/dr$ at $r=r_s$ as, 
$(dv_r/dr)_s=a\pm(a^2-b)^{1/2}$
which are the solutions of the equation 
$(dv_r/dr)_s^2-2a(dv_r/dr)_s+b=0$
where
\begin{eqnarray}
a&=&
	\frac{
        	\left(\frac{\partial\mathcal{N}}{\partial y}\right)_s
        	+\left(\frac{\partial\mathcal{N}}{\partial w}\right)_s
			\left(\frac{\partial a_s}{\partial v_r}\right)_s
		-\left(\frac{\partial\mathcal{D}}{\partial x}\right)_s
        	-\left(\frac{\partial\mathcal{D}}{\partial z}\right)_s
			\left(\frac{\partial \ell}{\partial r}\right)_s
        	-\left(\frac{\partial\mathcal{D}}{\partial w}\right)_s
			\left(\frac{\partial a_s}{\partial r}\right)_s
        }
        {	2\left[
        	\left(\frac{\partial\mathcal{D}}{\partial y}\right)_s
        	+\left(\frac{\partial\mathcal{D}}
			{\partial w}\right)_s\left(\frac{\partial a_s}{\partial v_r}\right)_s
                \right]
        },\nonumber\\
&& \\
b&=&\frac{
		-\left(\frac{\partial\mathcal{N}}{\partial x}\right)_s
        	-\left(\frac{\partial\mathcal{N}}{\partial z}\right)_s
			\left(\frac{\partial \ell}{\partial r}\right)_s
        	-\left(\frac{\partial\mathcal{N}}{\partial w}\right)_s
			\left(\frac{\partial a_s}{\partial r}\right)_s
	}
	{\left(\frac{\partial\mathcal{D}}{\partial y}\right)_s
        	+\left(\frac{\partial\mathcal{D}}
			{\partial w}\right)_s\left(\frac{\partial a_s}{\partial v_r}\right)_s}. 
\end{eqnarray}

In addition to the boundary conditions $\mathcal{D}=\mathcal{N}=0$, 
when the causal viscosity prescription is adopted, the accretion flows smoothly 
pass the viscous radius, $r_\nu$, where the condition $\mathcal{D}_\nu=0$ is satisfied. 
In this case, 
the two boundary conditions at the viscous radius, $r_\nu$, are given as 
$\mathcal{D}_\nu=\mathcal{N}=0$ at $r=r_\nu$. 
The method presented in this study uses the boundary conditions 
at the sonic radius, $r_s$, and the viscous radius, $r_\nu$, 
and do not use the boundary conditions in the outer region $r>r_\nu$ and 
the inner region $r_h<r<r_s$, where $r_h$ is the radius for the event horizon.  
%
%
\section{Calculation Method and Sample Solutions}

Based on the formula presented in the previous sections, we can 
calculate the global transonic solutions for ADAFs with the causal viscosity. 
The calculation procedures to obtain the global transonic solutions for ADAFs are as follows: 
\begin{enumerate}
\item First, we tentatively choose some value of $a_{s,s}$ for given values of $r_s$ and $j$, and  
calculate $v_{r,s}$, $\ell_s$, $(dv_r/dr)_s$, $(d\ell/dr)_s$ and $(da_s/dr)_s$ by using 
Eqs. (\ref{eq:angmom2}), (\ref{eq:dvr}) and (\ref{eq:das}).  
\item Next, we solve the solutions in the range $r_s<r<r_\nu$. 
In order to do this, we solve Eqs. (\ref{eq:angmom2}), (\ref{eq:dvr}) and (\ref{eq:das}) from the 
sonic point to the viscous point by using, e.g., the Runge-Kutta algorithm. 
Usually, for the initially selected value of $a_{s,s}$, the calculated solution does not pass the 
viscous point where two boundary conditions $\mathcal{D}_\nu=\mathcal{N}_\nu=0$ are satisfied. 
In such case, we return to step 1 and again choose the different values of $a_{s,s}$ for 
given values of $r_s$ and $j$. 
After repeating these procedures, 
we can determine the value $a_{s,s}$ which gives the solution 
satisfying the boundary conditions $\mathcal{D}=\mathcal{N}=0$ at $r=r_s$ and 
$\mathcal{D}_\nu=\mathcal{N}_\nu=0$ at $r=r_\nu$. 
\item After solving the solutions in $r_s<r<r_\nu$, 
we solve Eqs. (\ref{eq:angmom2}), (\ref{eq:dvr}) and (\ref{eq:das}) 
in the range $r_\nu<r$ by using the values of $a_{s,s}$ for given values of 
$r_s$ and $j$ by using, e.g., the Runge-Kutta algorithm. 
\item Finally, we solve Eqs. (\ref{eq:angmom2}), (\ref{eq:dvr}) and (\ref{eq:das})
in the range $r_h<r<r_s$ by using the values of $a_{s,s}$ for given values of 
$r_s$ and $j$ by using, e.g., the Runge-Kutta algorithm. 
%
\end{enumerate}
The third step and the fourth step can be interchanged. 
By this procedure, the transonic solutions are obtained for given values of $r_s$ and $j$. 
At the third step, the physical values, such as $d\ell/dr$,  at the viscous radius $r=r_\nu$ 
are used if necessary to obtain the solutions. 
Usually, the numerical calculation of the coupled equations are naturally performed 
through the viscous point, and the calculations in the third steps are successively done 
after passing through the viscous radius. 
However, in some cases, the numerical calculation is stopped at the viscous radius. 
In such case, we evaluate the physical values at the viscous point by using 
the L'Hopital's rule which give the value $d\ell/dr$ at the viscous radius $r=r_\nu$, and 
then we calculate the coupled equations for the global solution from the viscous radius 
to the outward direction $r>r_\nu$. 
In the same way as the calculation of $(dv_r/dr)_s$ as described in the last section, 
the value $(d\ell/dr)_{r=r_\nu}$ is calculated as 
%
$(d\ell/dr)_{r=r_\nu}=a_\nu\pm(a_\nu^2-b_\nu)^{1/2}$
where 
%
\begin{eqnarray}
a_\nu&=&
	\frac{1}{2}\left\{
        	\left(\frac{\partial\mathcal{N_\nu}}{\partial z}\right)_\nu
        	+\left(\frac{\partial\mathcal{N_\nu}}
			{\partial w}\right)_\nu \left(\frac{\partial a_s}{\partial \ell}\right)_\nu 
			-\left(\frac{\partial a_s}{\partial r}\right)_\nu
			-\left(\frac{\partial\mathcal{D_\nu}}{\partial x}\right)_\nu
	\right.\nonumber\\
	&&\left.
        	-\left[
			\left(\frac{\partial\mathcal{D_\nu}}{\partial y}\right)_\nu
			+\left(\frac{\partial\mathcal{D_\nu}}{\partial w}\right)_\nu
				\left(\frac{\partial a_s}{\partial v_r}\right)_\nu
			\right]
			\frac{\mathcal{N_\nu}}{\mathcal{D_\nu}}
    \right\}\nonumber\\
    &&
    \Bigg/    	\left[
        	\left(\frac{\partial\mathcal{D_\nu}}{\partial z}\right)_\nu
        	+\left(\frac{\partial\mathcal{D_\nu}}
			{\partial w}\right)_\nu \left(\frac{\partial a_s}{\partial \ell}\right)_\nu
                \right]
        ,\\
b_\nu&=&
	-\left\{
		\left(\frac{\partial\mathcal{N_\nu}}{\partial x}\right)_\nu
		+\left(\frac{\partial\mathcal{N_\nu}}{\partial w}\right)_\nu
			\left(\frac{\partial a_s}{\partial r}\right)_\nu
      \right.\nonumber\\
      &&\left.
			+\left[
			\left(\frac{\partial\mathcal{N_\nu}}{\partial y}\right)_\nu
			+\left(\frac{\partial\mathcal{N_\nu}}{\partial w}\right)_\nu
				\left(\frac{\partial a_s}{\partial v_r}\right)_\nu
			\right]
			\frac{\mathcal{N_\nu}}{\mathcal{D_\nu}}
	\right\}
	\nonumber\\
	&&
	\Bigg/
	\left[
	\left(\frac{\partial\mathcal{D_\nu}}{\partial z}\right)_\nu
        	+\left(\frac{\partial\mathcal{D_\nu}}
			{\partial w}\right)_\nu \left(\frac{\partial a_s}{\partial \ell}\right)_\nu
	\right]. 
\end{eqnarray}
Here, 
\begin{eqnarray}
\left(\frac{\partial\mathcal{D}_\nu}{\partial x}\right)_\nu&=&
\left(\frac{\partial\mathcal{D}_\nu}{\partial z}\right)_\nu=0
	,~~~
\left(\frac{\partial\mathcal{D}_\nu}{\partial y}\right)_\nu=\frac{2}{\alpha^{1/2}a_s}
	,~~~
\left(\frac{\partial\mathcal{D}_\nu}{\partial w}\right)_\nu=\frac{2}{a_s}
	,\nonumber \\
\left(\frac{\partial\mathcal{N}_\nu}{\partial x}\right)_\nu&=&
	-\frac{\mathcal{N}_\nu}{r}-\frac{(\ell-j)}{rv_r}\frac{d(\Omega_K r)}{dr}
	,\nonumber \\
\left(\frac{\partial\mathcal{N}_\nu}{\partial y}\right)_\nu&=&
	\frac{(\ell-j)}{rv_r} \left[\frac{v_r(\Omega_K r-2v_r)}{\alpha a_s^2}-2\right]
	,\nonumber \\
\left(\frac{\partial\mathcal{N}_\nu}{\partial z}\right)_\nu&=&
	\frac{1}{r} \left[\frac{v_r(\Omega_K r-2v_r)}{\alpha a_s^2}+2\right]
	,\nonumber \\
\left(\frac{\partial\mathcal{N}_\nu}{\partial w}\right)_\nu&=&
	\frac{-2(\ell-j)(\Omega_Kr-2v_r)}{rv_ra_s}
	,\nonumber \\
\left(\frac{\partial a_s}{\partial r}\right)&=& 
	\left(\frac{\gamma-1}{\gamma+1}\right)
	\left[
		a_s\frac{d\ln (\Omega_K/r)}{dr}-\frac{2\ell(\ell-j)}{a_s r^3}
	\right]
	,\nonumber \\
\left(\frac{\partial a_s}{\partial v_r}\right)&=& 
	-\frac{a_s}{v_r}\left(\frac{\gamma-1}{\gamma+1}\right)
	,~~~
\left(\frac{\partial a_s}{\partial \ell}\right)= 
	\left(\frac{\gamma-1}{\gamma+1}\right)\frac{(\ell-j)}{a_s r^2}
	,
\end{eqnarray}
where $d(\Omega_K r)/dr=-\Omega_K(r+2)/[2(r-2)]$. 
The values of $v_r$, $\ell$ and $a_s$ at the viscous radius are determined to 
satisfy $\mathcal{D}_\nu=\mathcal{N}_\nu=0$. 

It is noted that by using the method described above we do not use 
the outer boundary conditions for 
$r>r_\nu$ and the inner boundary conditions for $r<r_s$. 
If we want to the solutions for specified outer or inner boundary conditions, we should choose the 
values of $r_s$ and $j$ to satisfy the specified outer or inner boundary conditions. 
The correspondences between the values $r_s$ and $j$ and the outer solutions are investigated by 
Lu, Gu and Yuan (1999) for the special solutions satisfying the condition $\ell=j$ at $r=r_s$, 
and 
they provide solutions for the ADAF-thick disk solution, the ADAF-thin disk solution, and 
the alpha-type solution connecting the inner of outer regions. 
The special condition $\ell=j$ at $r=r_s$ simplify the equations described in the last section. 
In the case of the solutions including the cases for $\ell\neq j$ at $r=r_s$ 
which can be calculated by the method described above, 
we expect the basic picture is same as the solutions with $\ell=j$ at $r=r_s$ 
in Lu, Gu \& Yuan (1999). 
%
%

\begin{figure}
\centering
\includegraphics[width=0.45\textwidth]{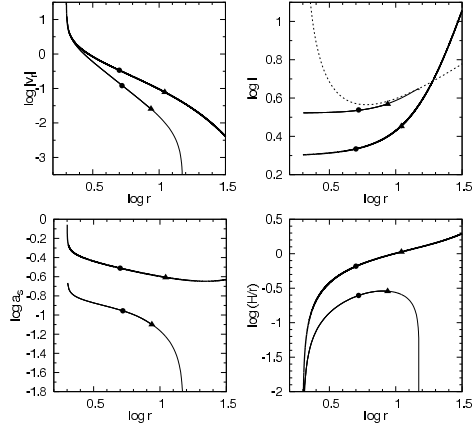}
\caption{
The sample solutions for ADAF-thick disk ({\it thick lines}) and 
ADAF-thin disk ({\it thin lines}) by solid lines for $\alpha=0.1$ and $\gamma=1.5$; 
the radial velocity $v_r$ ({\it top-left panel}), the angular momentum ({\it top-right panel}), 
the sound speed $a_s$ ({\it bottom panel }) and $H/r$ ({\it bottom-right panel}). 
The filled circles and triangles denote the positions of the sonic radius, $r_s$, 
and the viscous radius, 
$r_\nu$. 
%
%
%
%
In the panel of $\log \ell$, we also plot the line of 
the Keplerian angular momentum $\ell_K(=\Omega_K r^2)$ by a 
dotted line. 
%
\label{fig:ADAF-thick}}
\end{figure}

In Fig. \ref{fig:ADAF-thick}, 
we present the sample solutions for ADAF-thick disk ({\it thick lines}) 
and ADAF-thin disk ({\it thin lines}) by solid lines for $\alpha=0.1$ and $\gamma=1.5$; 
the radial velocity $v_r$ ({\it top-left panel}), the angular momentum ({\it top-right panel}), 
the sound speed $a_s$ ({\it bottom-left panel }) and $H/r$ ({\it bottom-right panel}).  
The filled circles and triangles denote the positions of the sonic radius, 
$r_s$, and the viscous radius, 
$r_\nu$ for all panels. 
We give the solutions in the range $r_h<r<10^{1.5}$, and the adopted values of 
$\log r_s$ and $j$ are $\log r_s=0.7$ and $j=2.0$. 
%
%
Here, we adopt the Runge-Kutta-Fehlberg algorithm for numerical calculations 
of the coupled differential equations of $v_r$, $\ell$ and $a_s$.  
%
%
The line of $\log|v_r|$ crosses the line of $\log [2\gamma/(\gamma+1)]^{1/2}a_s$ 
at the sonic radius, $r_s$, 
and the line of $\log c_\nu$ at the viscous radius, $r_\nu$, respectively. 
%
%
In the outer region $r>r_\nu$, the angular momentum of the solution achieves the Keplerian value.  
According to the all parameter search of the ADAF solutions by Lu, Gu \& Yuan (1999), 
the ADAF solutions are classified into the ADAF-thick disk solution, the ADAF-thin disk solution 
and the $\alpha$-type solution connecting the inner region or the outer region. 
While the parameter spaces of $r_s$ and $j$ 
for the ADAF-thick solutions and the $\alpha$-type solutions are
relatively wide, which are easily confirmed numerically, 
the parameter spaces for the ADAF-thin disk solutions are quite limited. 
By the calculation procedures presented in this paper, 
we can cover the all parameter spaces of the transonic solutions of ADAFs.   

\section{Discussion}
The formula in the present study become the formula for the acausal viscous prescription 
if we set the diffusion timescale $\tau_\nu$ to be null. 
So, the analytic expansion around the singular points presented in this paper 
are also used to calculate the global solution for ADAF with the acausal viscosity. 
In the case of the acausal viscosity, since the boundary condition at the viscous point 
can not be used, the boundary condition at the other radius is required, e.g. no-torque 
condition at the horizon. 
Although the finite diffusion velocity is physically motivated, 
the resultant solutions have generally non-zero torque at the horizon which seems to be  
incorrect from the point of view of the relativity. 
Also, in the causal viscous prescription, 
the differential equation for the angular momentum become more complex 
than that of the acausal viscosity.  
So, although the finite diffusion speed is more reasonable 
than the infinite diffusion speed, 
it is noted that 
the calculations of the ADAF structure by using the causal viscous prescription may not be 
clearly better than the calculations by using the acausal viscosity.  
In the fully general relativistic calculations performed by Gammie \& Popham (1998) which 
use the causal viscosity prescription, 
the boundary condition at the viscous point is used when calculating the global solution. 
In equation (60) of Gammie \& Popham (1998), 
although the viscous stress tensor 
at the horizon is finite if measured in the local rest frame, i.e. $t_{(r)(\phi)}\neq 0$, 
the viscous stress tensor 
at the horizon become null measured in the Boyer-Lindquist frame, i.e. $t^r_{~\phi}=0$,
because of $\mathcal{D}=0$ at the horizon. 
Since the location of the horizon is not the special location from the point of view of the local 
observer moving along the fluid's motion,  
non-zero stress in the local rest frame seems to be quite natural. 
The zero-torque condition arises only when we use the coordinate system having the coordinate 
singularity at the horizon such as the Boyer-Lindquist coordinate and measure the viscous stress 
in such frames. 
Then, the problem of the non-zero torque or zero-torque at the horizon is the issues 
which require the fully relativistic treatments with the spacetime structure 
and the frame-transformation between the observers measuring the viscous stress. 
So, it may be impossible to clearly resolve the torque problem at the horizon 
by using the calculations with the pseudo-Newtonian potential like the present study. 
%

\section{Concluding Remarks}
We present the basic equations and sample solutions for 
the steady-state global solutions of the ADAFs using the 
causal viscous prescription and the Paczy\'{n}skii-Wiita potential 
by the algorithm of explicit numerical integrations, such as the Runge-Kutta method. 
The calculation procedures to obtain these solutions are also presented. 
In this calculation procedure, we first solve the physical values at the sonic radius 
where L'Hopital's rule is used. 
The method presented in this paper enables us to stably solve the global solutions of 
ADAFs by the Runge-Kutta method, and the all parameter spaces of $r_s$ and $j$ for the 
transonic solutions of ADAFs can be covered in this method. 
If we set the diffusion timescale to be null, 
the formalism in this study includes the case of the acausal viscosity which is usually 
used in the past study. 
Since the calculation methods in this study use the analytic expansion around the singular 
point and the numerical integration are performed by the explicit integration, 
the numerical calculations become faster. 

\begin{acknowledgements}
The author is grateful to 
S. Mineshige and K. Watarai for useful discussion. 
This research was partially supported by the Ministry of Education,
Culture, Sports, Science and Technology, Grant-in-Aid for 
Japan Society for the Promotion of Science (JSPS) Fellows (17010519).
\end{acknowledgements}

\end{document}